\documentclass[pra,a4paper,twocolumn,showpacs,superscriptaddress]{revtex4}

\usepackage{mathrsfs}
\usepackage{amsmath}
\usepackage{amssymb}
\usepackage{graphicx}
\usepackage{color}

\newcommand{\tr}{\operatorname{tr}}

\begin{document}
\bibliographystyle{apsrev}

\title{Entanglement in a simple quantum phase transition}

\author{Tobias J.\ Osborne}
\email[]{osborne@physics.uq.edu.au}
\affiliation{Department of Mathematics, University of Queensland 4072,
Australia}
\affiliation{Centre for Quantum Computer Technology and Department of Physics,
University of Queensland 4072, Australia}
\author{Michael A.\ Nielsen}
\email[]{nielsen@physics.uq.edu.au}
\affiliation{Centre for Quantum Computer Technology and Department of Physics,
University of Queensland 4072, Australia}

\date{\today}

\begin{abstract} What entanglement is present in naturally occurring
physical systems at thermal equilibrium?  Most such systems are
intractable and it is desirable to study simple but realistic
systems which can be solved. An example of such a system is the
$1D$ infinite-lattice anisotropic $XY$ model.  This model is
exactly solvable using the Jordan-Wigner transform, and it is
possible to calculate the two-site reduced density matrix for all
pairs of sites. Using the two-site density matrix, the
entanglement of formation between any two sites is calculated for
all parameter values and temperatures. We also study the
entanglement in the transverse Ising model, a special case of the
$XY$ model, which exhibits a quantum phase transition.  It is
found that the next-nearest neighbour entanglement (though not the
nearest-neighbour entanglement) is a maximum at the critical
point.  Furthermore, we show that the critical point in the
transverse Ising model corresponds to a transition in the
behaviour of the entanglement between a single site and the
remainder of the lattice.
\end{abstract}

\pacs{03.65.Ud, 73.43.Nq, 05.50.+q}
\maketitle

\section{Introduction}\label{sec:intro}

It seems to be a truism in quantum physics that strongly entangled systems 
exhibit complicated behaviour which is 
difficult to quantify.  Two practical examples of this `principle' 
are the conventional superconductor 
\cite{schrieffer:1964a, tinkham:1996a} and the 
fractional quantum Hall effect 
(FQHE) \cite{prange:1990a}. 
In both cases, for certain parameter regimes, the system enters
a very interesting \emph{entangled} state (the BCS ground state for the
superconductor \cite{bardeen:1957a, bardeen:1957b}, 
and the Laughlin ground state for the FQHE \cite{laughlin:1983a}).  
For many years 
these systems resisted attempts to understand them using 
reasoning based on classical methods
\cite{endnote60}.
It required a major breakthrough,
the construction of an insightful ground state ansatz, to elucidate 
the physics of both the FQHE and the superconductor.  The key feature of
both systems, which makes it hard to explain them classically, 
appears to be that their ground states are \emph{strongly entangled}.

Entanglement is a uniquely quantum property of any
\emph{nonlocal} superposition-state of two or more quantum systems
\cite{schrodinger:1935a, bell:1964a, bennett:1996a}.  
Such states are typified by the Bell state $|\Psi^-\rangle =
\frac{1}{\sqrt2}(|01\rangle-|10\rangle)$.  The many curious
features of entangled states  
have motivated considerable research.  A
remarkable consequence of this work is the emerging understanding of
entanglement as a \emph{resource} \cite{bennett:1996b, bennett:1996a},
like energy, which can be used to accomplish interesting physical
tasks.  

The similarities between entanglement and energy appear to be more
than just superficial.  It turns out to be possible to \emph{quantify}
the entanglement present in a given quantum state.  This allows the
development of quantitative high-level principles governing the
behaviour of entangled states, independent of their particular
physical representation.  These principles can be seen as analogous to
the laws of thermodynamics governing the behaviour of energy,
independent of the specific form in which it is given to us.  We hope
that the quantitative theory of entanglement may provide a powerful
unifying framework for the understanding of \emph{complex} quantum
systems.  This is because, when viewed in terms of their entanglement
content, a large number of apparently different states turn out to be
equivalent.

This paper is one step in testing the hypothesis \cite{nielsen:1998a,
preskill:2000a, osborne:2001a, zanardi:2002a} that the study of
complex quantum systems may be simplified by first analysing the
static and dynamic entanglement present in those systems.  We will
attempt to perform such an analysis in a representative system chosen
from condensed matter physics, specifically, the $XY$ 
model~\cite{lieb:1961a}.  The
signature of complexity in this system is the occurrence of a quantum
phase transition.
	
Quantum phase transitions (QPT) are a qualitative change in the ground
state of a quantum many-body system as some parameter is varied
\cite{sachdev:1999a, sondhi:1997a}.  Unlike ordinary phase
transitions, which occur at a nonzero temperature, the fluctuations in
a QPT are fully quantum. Typically, at the \emph{critical point} in
parameter space where a QPT takes place, long-range correlations in
the ground state also develop.  The existence of a QPT in a quantum
many-body system strongly influences the behaviour of the system near
the critical point, with the development of long-range correlations
and a nonzero expectation value for an order parameter
\cite{sachdev:1999a}.

In \cite{osborne:2001a} it was argued that QPTs are genuinely \emph{quantum
mechanical} in the sense that the property responsible for the long-range 
correlations is entanglement.  It was also argued that the system state is
strongly entangled at the critical point.  It would be desirable, to begin
with, to show that systems near quantum critical points can be simply
characterised in terms of their entanglement content.
Unfortunately, such a proof seems very difficult.  We need first to
understand the entanglement in such systems before proposing a
classification scheme based on entanglement content.  At the moment the most
promising technique to study entanglement in critical quantum systems 
appears to be the renormalization 
group, which is the standard way to obtain information about systems
at and near criticality.

The renormalization group (RG) is based on the notion that physics at
small length scales (and hence higher energy scales) should not affect
physics at much larger length scales.  The RG is, in fact, a family of
methods which can be applied to learn non-perturbative information
about strongly interacting systems.  The development of the
renormalization group (see, for example, \cite{fisher:1998a,
cardy:1996a} for a review) has shown that phase transitions are
\emph{universal} in the sense that many properties of the system do
not depend on the detailed dynamics of the system under
consideration. Instead, using RG techniques, it has been shown that
phase transitions depend only on certain global properties, such as
symmetry and dimension. We would like to apply the ideas of the RG to
calculate entanglement quantities in systems exhibiting a quantum
phase transition.  To see if this is possible, it is desirable to
first carry out
exact calculations in order to determine if similar universality
properties govern the entanglement present in such systems.  The
purpose of this paper is therefore to do such calculations for the
$XY$ model.

Unfortunately the modern theory of entanglement (see, for example, the
review issue of {\em Quantum Information and Computation}
\cite{horodecki:2001a,wootters:2001a,horodecki:2001b,nielsen:2001a})
is only partially developed, and at the present time can only be
applied in a limited number of scenarios. In these limited scenarios
well-developed analytic tools exist to quantify the structure of
entanglement present in a system.  Two important scenarios are (a) the
case of a pure state of a bipartite system, that is, a system
consisting of only two components; and (b) a mixed state of two
spin-$\frac 12$ particles.

For this reason, we focus our
investigation on two types of calculation for the $XY$ model. The
first calculation is of the entanglement between a single site in
the lattice and the rest of the system, for the ground state of the model.  
The
second calculation is of the entanglement between two sites of the
lattice at arbitrary temperatures and separations, allowing us to
determine whether there are truly quantum features present in the
two-body correlations in the system.  Thus, although we do not
obtain an understanding of the three- and more-party entanglement
present in the system, we do calculate significant partial
information characterising the entanglement.

The entanglement present in condensed-matter systems has been
investigated previously by a number of authors
\cite{nielsen:1998a, wootters:2000a, arnesen:2001a,
wang:2001a, meyer:2001a, gunlycke:2001a, wang:2001b, wang:2001d, fu:2002a,
zanardi:2002a, wang:2002a}. It was considered by Nielsen
\cite{nielsen:1998a} who studied the Heisenberg model on two sites
analytically. An expression for the ground state entanglement in
the infinite $1D$ Heisenberg chain was obtained soon after by
Wootters \cite{wootters:2000a}. 
Numerical
calculations of entanglement in the Heisenberg model on a small
number of sites were carried out by Arnesen \emph{et al}.\
\cite{arnesen:2001a}. Arnesen \emph{et al}.\ identified parameter
regions where there is appreciable \emph{thermal entanglement},
which is entanglement present at nonzero temperatures.  Recent
studies include the numerical calculation of entanglement in the
transverse Ising model on small numbers of sites
\cite{gunlycke:2001a}, and analytic computations of entanglement in
the $XY$ model on $2$ sites \cite{wang:2001a} and $3$ sites
\cite{wang:2001b}.  Additional studies have been carried out on itinerant
fermion systems \cite{zanardi:2002a} 
and other small condensed matter systems related to the
$XY$ model \cite{wang:2001d, wang:2001a, fu:2002a, wang:2002a}.

The structure of this paper is as follows.  In Sec.~\ref{sec:exsol}
the exact solution and calculation of the correlation functions for
the $XY$ model is outlined using the Jordan-Wigner transform.  The
thermal ground state properties of this system are considered in
Sec.~\ref{sec:gseis}, focusing on the special case of the transverse
Ising model, and the role entanglement plays in the quantum phase
transition in this model. Thermal entanglement
in the transverse Ising model is then calculated in
Sec.~\ref{sec:2sent}.  We conclude in Sec.~\ref{sec:conc}, and sketch
some possible future research directions.

\section{Exact solution of the $XY$ model}\label{sec:exsol}

In this section we consider the exact solution of the $XY$ model
on $N$ sites, which is facilitated by use of the Jordan-Wigner
transform \cite{jordan:1928a}.  The observables that are important
for the calculation of the entanglement are evaluated in the
\emph{large-$N$} or \emph{thermodynamic} limit. The two
fundamental objects constructed in this study are the one- and
two-site density matrices.  From knowledge of these matrices it is
possible to calculate the one- and two-party entanglement
occurring in the $XY$ model. The solution of the $XY$ model is
well known, and the procedure outlined in this section to solve it
follows the standard method \cite{lieb:1961a, barouch:1970a,
chakrabarti:1996a, sachdev:1999a}.  The main result in this 
section is the explicit construction of the one- and two-party density
matrices for the $XY$ model at thermal equilibrium.

The Hamiltonian for the anisotropic $XY$ model on a $1D$ lattice with
$N$ sites in a transverse field is given by \cite{anderson:1958a}
\begin{equation}\label{eq:xyham} H = -\sum_{j=0}^{N-1}\left(
\frac{\lambda}{2}
\left[(1+\gamma)\sigma^x_j\sigma^x_{j+1}+(1-\gamma)\sigma^y_j\sigma^y_{j+1}
\right]+\sigma^z_j\right), \end{equation} where $\sigma^a_j$ is the
$a$th Pauli matrix ($a = x,y$ or $z$) at site $j$, $\gamma$ is the
degree of anisotropy, and $\lambda$ is the inverse strength of the
external field.  We assume cyclic boundary conditions, so that the $N$th
site is identified with the $0$th site. The standard procedure used to solve
Eq.~(\ref{eq:xyham}) is to transform the spin operators $\sigma^a_j$ into
fermionic operators via the Jordan-Wigner transform
\begin{align}\label{eq:fermop1} c_i&\equiv\prod_{j=0}^{i-1}
[-\sigma_j^z]\sigma_i^-, \\ \label{eq:fermop2}
c_i^\dag &= \prod_{j=0}^{i-1}
[-\sigma_j^z] \sigma_i^+, 
\end{align} 
where 
\begin{equation}
\sigma_i^+\equiv\frac{1}{2}(\sigma_i^x+i\sigma_i^y), \quad
\sigma_i^-\equiv\frac{1}{2}(\sigma_i^x-i\sigma_i^y). 
\end{equation}
It is easy to verify that $c_i$ satisfy the fermionic anticommutation
relations
\begin{equation} 
\{c_i,c_j^\dag\}=\delta_{ij}, \quad
\{c_i,c_j\}=0.  
\end{equation} 
In terms of the fermionic operators,
Eqs.~(\ref{eq:fermop1})-(\ref{eq:fermop2}), 
the Hamiltonian Eq.~(\ref{eq:xyham}) assumes the
quadratic form
\begin{equation}\label{eq:fermham} H =
\left(\sum_{i,j=0}^{N-1}c_i^\dag
A_{i,j}c_j+\frac{1}{2}\sum_{i,j=0}^{N-1}\left(c_i^\dag B_{i,j}
c_j^\dag+\text{h.c.}\right)\right)+N,
\end{equation}
where $A_{i,i} =
-1$, $A_{i,i+1}=-\frac{1}{2}\gamma\lambda=A_{i+1,i}$,
$B_{i,i+1}=-\frac{1}{2}\gamma\lambda$,
$B_{i+1,i}=\frac{1}{2}\gamma\lambda$ and all the other $A_{i,j}$ and
$B_{i,j}$ are zero.  The quadratic Hamiltonian Eq.~(\ref{eq:fermham}) may
be diagonalised by making a linear transformation of the fermionic
operators \begin{align}\label{eq:newop} \eta_q &=
\sum_{i=0}^{N-1}\left( g_{qi}c_i+h_{qi}c_i^\dag\right),\\
\eta_q^\dagger &= \sum_{i=0}^{N-1}\left(
g_{qi}c_i^\dag+h_{qi}c_i\right), \end{align} where $q =
-N/2,-N/2+1,\ldots,N/2-1$ and the $g_{qi}$ and $h_{qi}$ can be chosen
to be real.  By requiring that the operators $\eta_q$ obey fermionic
anticommutation relations,
and that the Hamiltonian Eq.~(\ref{eq:xyham}) be manifestly 
diagonal when expressed in terms of the fermionic modes $\eta_q$, 
the following two coupled matrix equations must hold 
\begin{align} 
(A-B)\Phi_q &= \omega_q\Psi_q, \\
(A+B)\Psi_q &= \omega_q\Phi_q, \end{align} where the components of the
two column vectors $\Phi_q$ and $\Psi_q$ are given by \begin{align}
[\Phi_q]_i &= g_{qi}+h_{qi},\\ [\Psi_q]_i &= g_{qi}-h_{qi}.
\end{align} 
The quadratic Hamiltonian Eq.~(\ref{eq:fermham}), when expressed
in terms of the operators $\eta_q$, takes the
diagonal form 
\begin{equation} 
H =
2\sum_q\omega_q\eta_q^\dag\eta_q-\sum_q\omega_q, 
\end{equation} 
where
\begin{equation}
\omega_q=\sqrt{(\gamma\lambda\sin \phi_q)^2 + (1+\lambda\cos \phi_q)^2},
\end{equation}
and $\phi_q=2\pi q/N$.

Now that the $XY$ Hamiltonian has been diagonalised we can
calculate the one- and two-site density matrices. 
Much of the remainder of this paper is concerned with the case 
where the system is at thermal equilibrium at 
temperature $T$.
The density matrix for the $XY$ model at thermal equilibrium
is given by the canonical ensemble $\rho =
e^{-\beta H}/\mathcal{Z}$, where $\beta \equiv 1/k_BT$, and
$\mathcal{Z}=\tr(e^{-\beta H})$ is the partition function.  The thermal
density matrix is diagonal when expressed in terms of the Jordan-Wigner
fermionic operators $\eta_q$. 
Our
interest lies in calculating the quantum correlations present in the system
as a function of the parameters $\beta$, $\gamma$, 
$\lambda$.  
In general this problem requires
knowledge of all the possible spin correlation functions.  These correlators
are typically very difficult to calculate from $\rho$ as it is diagonal in
terms of the $\eta_q$'s, which are complicated nonlocal functions
of the original spin operators.  Fortunately, the only correlation
functions which we require are the one- and two-point correlation
functions.  The evaluation of these functions has been carried
out previously \cite{barouch:1970a, barouch:1971a}.

The one- and two-site density matrices may be constructed from the one- and
two-point correlation functions, using the
\emph{operator expansion} for the
density matrix of a system of $N$ spin-$\frac12$ particles
in terms of tensor products of Pauli matrices.
For the single-site density matrix $\rho_1$ for the first spin  --- equal,
by translational symmetry, to the state $\rho_i$ of a single spin at
an arbitrary site ---
the operator expansion reads
\begin{equation}
\rho_1 =\tr_{\hat{i}}(\rho)=
\frac{\sum_{\alpha=0}^{3}q_\alpha\sigma^\alpha_i}{2},
\end{equation}
where $\tr_{\hat{i}}$ is the partial
trace over all degrees of freedom except the spin at site $i$, 
$\sigma_i^\alpha$ are the Pauli matrices
acting on the site $i$ with the convention $\sigma_i^0=I_i$, and
the coefficients $q_{\alpha}$ are real.
The coefficients
$q_\alpha$ are determined by the relation 
\begin{equation}\label{eq:ssope}
q_\alpha = \tr(\sigma^i_\alpha\rho) = \langle\sigma^i_\alpha\rangle.
\end{equation}

To completely specify the single-site density matrix requires knowledge of
three expectation values
($q_0=1$ because $\rho_1$ must have trace unity).  However, because the
Hamiltonian for the $XY$ model Eq.~(\ref{eq:xyham}) 
possesses symmetries
it is possible to reduce this number to one.  First of all, the 
Hamiltonian is real, so that $\rho_1^* = \rho_1$.  As the matrix $\sigma^y$
is imaginary this means that $q_2$ must be zero.  The second symmetry
that the $XY$ Hamiltonian possesses is the global phase flip symmetry 
\begin{equation}\label{eq:glsf}
U_{\text{PF}} =\prod_{j=0}^{N-1} \sigma^z_j.
\end{equation}
This symmetry implies that $[\sigma^z, \rho_1]=0$, so 
forcing  
$q_3$ to be zero.  The single-site density matrix $\rho_1$
is therefore determined solely by $q_1$.

For the two-site density matrix, which is the joint state of
two spins at sites $i$ and $j$, the
operator expansion takes the form
\begin{equation}\label{eq:ope} \rho_{ij} =
\tr_{\widehat{ij}}(\rho) = 
\frac{\sum_{\alpha,\beta=0}^3p_{\alpha\beta}\sigma_i^\alpha\otimes\sigma_j^\beta}{4}.
\end{equation} 
The coefficients are determined by the relation 
\begin{equation}
p_{\alpha\beta} = \tr(\sigma_i^\alpha\sigma_j^\beta\rho_{ij}) =
\langle\sigma_i^\alpha\sigma_j^\beta\rangle, 
\end{equation} 
so
that if the relevant correlation functions are known it is
possible to construct the two-site density matrix completely.

The operator expansion Eq.~(\ref{eq:ope}) implies that we need sixteen
correlation functions to construct the two-site density matrix.  However,
as in the case of the single-site density matrix, 
this number can be reduced by appealing to the symmetries of the 
the Hamiltonian.  
Translational invariance of the lattice means that the density matrix
depends only on the distance $r=|j-i|$ between the spins,
that is, $\rho_{ij}=\rho_{0r}$.  Reflection symmetry about any site also
means that $\rho_{ij}=\rho_{ji}$.  Also, since the Hamiltonian
is real, ${\rho_{ij}}^*=\rho_{ij}$.  Finally, the the global phase flip
symmetry implies that $[\sigma^z_i\sigma^z_j,\rho_{ij}]=0$.
The symmetries of the $XY$ model require that the only nonzero
coefficients in the operator expansion Eq.~(\ref{eq:ope}) are
$p_{00}$, $p_{03}$, $p_{30}$, $p_{11}$, $p_{22}$, and $p_{33}$.
Furthermore, $p_{00}=1$ because the density matrix must have trace unity,
and $p_{03}=p_{30}$.

In the thermodynamic limit, $N\rightarrow\infty$, sums that appear
in the expectation values are replaced by integrals, and the
correlation functions for the $XY$ model can be reduced to
quadratures \cite{lieb:1961a, pfeuty:1970a, barouch:1970a,
barouch:1971a}.  The calculations are rather involved, and we
merely summarise the results here.  In thermal equilibrium, for
arbitrary $\gamma$ and $\lambda$, the transverse magnetisation
$\langle\sigma^z\rangle$ is given by \cite{barouch:1970a}
\begin{equation}\label{eq:trmag} \langle\sigma^z\rangle =
-\frac{1}{\pi}\int_0^\pi d\phi\,
(1+\lambda\cos\phi)\frac{\tanh(\frac{1}{2}\beta\omega_\phi)}
{\omega_\phi}, \end{equation} 
where we abuse notation and write
$\omega_\phi \equiv \omega_q$ to indicate the replacement of
$\phi_q$ with the continuous variable $\phi$ which results from
the thermodynamic limit $\phi_q\rightarrow\phi$.  

The two-point
correlation functions are given by \cite{barouch:1971a}
\begin{align} \langle \sigma_0^x\sigma_r^x\rangle &=
\begin{vmatrix} G_{-1} & G_{-2} & \cdots & G_{-r} & \\ G_0 &
G_{-1} & \cdots & G_{-r+1} & \\ \vdots & \vdots & \ddots & \vdots
& \\ G_{r-2}& G_{r-3}& \cdots & G_{-1} &
\end{vmatrix}, \label{eq:corr1} \\ \langle \sigma_0^y\sigma_r^y\rangle
&= \begin{vmatrix} G_{1} & G_{0} & \cdots & G_{-r+2} & \\ G_2 & G_{1}
& \cdots & G_{-r+3} & \\ \vdots & \vdots & \ddots & \vdots & \\ G_{r}&
G_{r-1}& \cdots & G_{1} & \end{vmatrix}, \label{eq:corr2} \\ \langle
\sigma_0^z\sigma_r^z\rangle &= 4\langle\sigma^z\rangle^2 - G_rG_{-r}
\label{eq:corr3}, \end{align} where \begin{multline} G_r =
\frac{1}{\pi}\int_0^\pi d\phi\, \cos(\phi r)
(1+\lambda\cos\phi)\frac{\tanh(\frac{1}{2}\beta\omega_\phi)}{\omega_\phi} \\
-\frac{\gamma\lambda}{\pi}\int_0^\pi d\phi\, \sin(\phi
r)\sin(\phi)\frac{\tanh(\frac{1}{2}\beta\omega_\phi)}{\omega_\phi}.
\end{multline}

Summarising, in the thermodynamic limit we may write the 
single-site density matrix $\rho_1$ entirely
in terms of the transverse magnetisation Eq.~(\ref{eq:trmag}),
\begin{equation}
\rho_1=\frac{I+\langle\sigma^z\rangle\sigma^z}{2}.
\end{equation}
Similarly, the 
two-site density matrix $\rho_{0r}$ can be written entirely in
terms of the correlation functions Eq.~(\ref{eq:corr1}),
Eq.~(\ref{eq:corr2}), Eq.~(\ref{eq:corr3}) and the transverse
magnetisation,
\begin{equation}
\rho_{0r} = \frac{I_{0r} +
\langle\sigma^z\rangle(\sigma_0^z+\sigma_r^z) +
\sum_{k=1}^3\langle
\sigma_0^k\sigma_r^k\rangle\sigma^k_0\sigma^k_r} {4}.
\end{equation}

\section{Ground state entanglement for the transverse Ising
and $XY$ models}\label{sec:gseis}

In this section we discuss the quantum correlations occurring in the
ground state of lattice systems undergoing a quantum phase transition.
We argue that the critical point corresponds to the situation where
the lattice is \emph{critically entangled}, where, somewhat loosely,
we define critically entangled to mean that entanglement is present on
all length scales.  In subsection \ref{ssec:timgs} we outline the
properties of the ground state of the transverse Ising model, which is
a simple subclass of the anisotropic $XY$ model.  In subsection
\ref{ssec:gsent} the contribution to the ground-state correlations
from one- and two-party entanglement in the $XY$ model is calculated
explicitly in order to illustrate the sharp peak in the entanglement
at the critical point.  Finally, in subsection \ref{ssec:entshare} we
discuss how the properties of shared entanglement may be related to
critical quantum lattice systems.

In \cite{osborne:2001a} it was argued that the physical origin of the
correlations which occur in systems exhibiting a quantum phase
transition is quantum entanglement.  We reproduce the argument of
\cite{osborne:2001a} here in order that this study be self-contained.
For concreteness, we restrict our attention to a lattice of
spin-$\frac12$ particles.

Suppose the ground state of a quantum lattice system were not
entangled, that is, it is a product state.  Then a simple calculation
shows that the spin-spin correlation function
$\langle\sigma_i^\alpha\sigma^\beta_j \rangle - \langle\sigma_i^\alpha
\rangle \langle\sigma_j^\beta \rangle $ is identically zero.  Thus, if
the correlation function is non-zero then the ground state must be
entangled.  Furthermore, we conjecture that large values of the
correlation function imply a highly entangled ground state; it is
interesting open problem to prove a precise form of this conjecture.

For general quantum lattice systems the correlation function decays
exponentially as a function of the separation $|i-j|$ when the system
is far from criticality~\cite{sachdev:1999a}.  When the system is at a
critical point, the correlations decay only as a polynomial function
of the separation.  At this point a fundamental change in the ground
state has occurred.  

We believe that when a system approaches a critical point the
structure of the entanglement in the ground state undergoes a
transition.  Further, we conjecture that the nature of this transition
is governed by a change in the spatial extent of the entanglement.
The entanglement between a single spin and the rest of the lattice
away from the critical point must be bounded in finite regions because
the correlations are damped exponentially.  At the critical point
correlations develop on all length scales, and the physical property
responsible for these correlations, entanglement, should become
present at all length scales as well.  We believe that a fundamental
transition in the nature of the entanglement in the system occurs at
this point; in some sense, at the critical point the state is
delocalized, compared with the local nature of the entanglement away
from the critical point.  If this physical picture is correct, there
should be evidence of entanglement developing on all length scales in
the one- and two-party entanglement results.

As described in detail below, the ground state of the $XY$ model
exhibits the features we have described in the previous paragraphs.
That is, maximality of the entanglement at criticality, and evidence
that a transition in the entanglement structure takes place at the
critical point.  Although much work remains to be done to flesh out
this physical picture, we believe that further research will show that
these are generic properties of critical quantum systems.

\subsection{Properties of the transverse Ising model ground
state} \label{ssec:timgs}

The ground state of the $XY$ model is very complicated with many
different regimes of behaviour \cite{barouch:1970a,
barouch:1971a}. For the sake of clarity, we focus most of our
discussions on the transverse Ising model, which arises as the
zero-anisotropy limit $\gamma\rightarrow1$ in Eq.~(\ref{eq:xyham}).
The reason for this particular choice is because the transverse Ising
model is the simplest quantum lattice system to exhibit a quantum
phase transition \cite{sachdev:1999a}. The central goal in this
section is to illustrate the intimate relationship between the
entanglement structure of the ground state and the quantum phase
transition.  In particular, the calculations for the transverse Ising
model provide the clearest evidence for the conjecture that the
critical point corresponds to the situation where the lattice is most
entangled.

The Hamiltonian for the transverse Ising
model may be obtained from the $XY$ model Hamiltonian,
Eq.~(\ref{eq:xyham}), by setting $\gamma=1$:
\begin{equation}
H = -\sum_{j=0}^{N-1} \left( \lambda\sigma_j^x\sigma_{j+1}^x
+\sigma_j^z \right).
\end{equation}
The structure of the transverse Ising model ground state changes
dramatically as the parameter $\lambda$ is varied. The dependence
of the ground state on $\lambda$ is quite complicated.  However, it
is possible to investigate the $\lambda=0$ and
$\lambda\rightarrow\infty$ limits exactly.

When $\lambda$ approaches zero, the transverse Ising model ground
state becomes a product of spins pointing in the positive $z$
direction, \begin{equation}
|0\rangle_{\lambda\rightarrow0}\approx\cdots|\uparrow\rangle_j|\uparrow\rangle_{j+1}
\cdots.  \end{equation} In the $\lambda\rightarrow\infty$ limit the
ground state again approaches a product of spins pointing in the positive $x$
direction, \begin{equation}
|0^+\rangle_{\lambda\rightarrow\infty}\approx\cdots|\rightarrow\rangle_j|\rightarrow\rangle_{j+1}
\cdots.  \end{equation} The $\lambda\rightarrow\infty$ limit is
fundamentally different from the $\lambda=0$ case because the
corresponding ground state is doubly degenerate under the global phase
flip, Eq.~(\ref{eq:glsf}), where \begin{equation}
|0^-\rangle_{\lambda\rightarrow\infty}\equiv
U_{\text{PF}}|0^+\rangle_{\lambda\rightarrow\infty}\approx\cdots|\leftarrow\rangle_j|\leftarrow\rangle_{j+1}
\cdots \end{equation} is a second ground state.  The $\lambda=0$
ground state is invariant under the global phase flip.  We note that
in both limits the ground state approaches a product state.

Using the solutions obtained for the limiting cases of $\lambda$ we
can qualitatively describe the ground state as $\lambda$ is
varied. When $\lambda$ is small, the exchange term
$\sigma_j^x\sigma_{j+1}^x$ may be regarded as a perturbation, and
perturbation theory may be used.  In this case the ground state
becomes a superposition of the unperturbed ground state and low-lying
excitations in such a way that the small-$\lambda$ ground state
remains invariant under the global phase flip.  

When $\lambda$ is much greater than one, $1/\lambda$ is a small
parameter and perturbation theory may again be used to show that the
now-degenerate ground states are a superposition of the unperturbed
ground states $|0^{+,-}\rangle$ and low-lying excitations.  The
degeneracy of the ground state under the global phase flip remains for
$\lambda$ large. (This degeneracy, along with the invariance of the
ground state $|0\rangle$ under $U_{\text{PF}}$ may be established
nonperturbatively \cite{sachdev:1999a}.) 

When $\lambda=1$ a fundamental transition in the form of the ground
state occurs. The symmetry under the global phase flip breaks at this
point and the system develops a nonzero magnetisation
$\langle\sigma^x\rangle\not=0$ which grows as $\lambda$ is increased.
The magnetisation is the \emph{order parameter} which identifies the
existence of a new phase.

Now that we have outlined the structure of the ground state for
the transverse Ising model as a function of $\lambda$ we have a
basic physical picture with which to interpret the exact results.

The calculation of the entanglement between a single site and the rest
of the lattice requires construction of the single-site density matrix
for the ground state.  While the single-site density matrix for the
thermal state was constructed in Sec.~\ref{sec:exsol}, there is a
distinction between the zero-temperature limit of the thermal density
matrix and the ground state, because of the possible ground-state
degeneracy.  In the following, when referring to the \emph{ground
state} of the system, we suppose the system to be in one of the
possible degenerate eigenstates $|0^+\rangle$ or $|0^-\rangle$ rather
than any other linear combination.  It does not matter which of the
two is chosen to be `the' ground state because all the entanglement
quantities calculated in this paper do not depend on the choice, due
to the local symmetry connecting the two states.  Therefore, without
loss of generality, when the system is in the ground state we choose
the system to be in the eigenstate $|0^+\rangle$ for $\lambda > 1$ and
$|0\rangle$ for $\lambda\le1$.  For simplicity we will identify
$|0^+\rangle$ with $|0\rangle$ when $\lambda$ is greater than or equal
to one.

The \emph{zero temperature state}, $\rho_0$, of the $XY$ model may be found by
taking the limit $\beta\rightarrow\infty$ of the canonical
ensemble,
\begin{equation}
\rho_0 = \lim_{\beta\rightarrow\infty}\frac{e^{-\beta
H}}{\mathcal{Z}}.
\end{equation}
When the ground state is nondegenerate the zero temperature state
is the same as the ground state of the system,
$\rho_0=|0\rangle\langle0|$.  However, if the ground state is
degenerate the zero temperature ensemble becomes an equal mixture of
 all the possible ground states.  For the transverse Ising model the
zero temperature state may be written
\begin{equation}\label{eq:gsens}
\rho_0=\frac12|0^+\rangle\langle0^+| +
\frac12|0^-\rangle\langle0^-|.
\end{equation}
In
order to differentiate between the actual ground state $|0\rangle$ of
the $XY$ model and the zero temperature ensemble we refer to
$\rho_0$ as the \emph{thermal ground state}.

In general, the canonical ensemble $\rho$ possesses the same
symmetries as the Hamiltonian Eq.~(\ref{eq:xyham}).  This is a simple
consequence of the identity $[U,H]=0$, where $U$ is some unitary or
antiunitary operator representing the symmetry operation.  The
invariance follows from $[U,\rho]=0$, so that $U\rho U^\dag=\rho$. In
particular, while each individual degenerate ground eigenstate may not
possess the same symmetries as the Hamiltonian, the thermal ground
state $\rho_0$ has all the same symmetries.

The quantum phase transition in the transverse Ising model separates
two different phases, the \emph{paramagnetic} phase where the
magnetisation $\langle\sigma^x\rangle$ is zero, and the
\emph{ferromagnetic} phase where the magnetisation becomes nonzero.
Associated with the development of a nonzero value for the order
parameter $\langle\sigma^x\rangle$ is the breaking of the phase flip
symmetry.  The symmetry breaking present in the ground state
$|0\rangle$ is a key feature of the quantum phase transition, and is
responsible for the development of non-zero order parameter $\langle
\sigma^x \rangle$ associated with the ferromagnetic phase. (In practice,
small external perturbations force spontaneous symmetry breaking of
the phase flip symmetry, and the system will choose one or the other
ground state, so this order parameter is, in principle, observable.)
This symmetry breaking cannot occur in the thermal ground state.  For
this reason, we will be most interested in properties of $|0\rangle$
rather that $\rho_0$.  For each of the degenerate ground eigenstates
$|0^+\rangle$ and $|0^-\rangle$ the global phase flip symmetry is
broken, so the terms which were set to zero in the operator expansion
Eq.~(\ref{eq:ope}), as a consequence of the symmetry
Eq.~(\ref{eq:glsf}), may become nonzero.

The single-site density matrix $\rho_1$ for the ground state of
the Ising model is obtained by taking a partial trace over all but
one site of
$|0\rangle\langle0|$. 
In general, because the global
phase-flip symmetry may be broken, the operator expansion for
$\rho_1$ is only  constrained by the reality condition
$\rho_1^*=\rho_1$.
Therefore, typically, two parameters are required to specify
$\rho_1$ completely, the magnetisation $\langle\sigma^x\rangle$
and the transverse magnetisation $\langle\sigma^z\rangle$:
\begin{equation}
\rho_1 = \frac{I+\langle\sigma^x\rangle \sigma^x +
\langle\sigma^z\rangle\sigma^z}{2}.
\end{equation}
It is difficult to calculate the magnetisation
$\langle\sigma^x\rangle$ of the ground state explicitly because
its expression in terms of Jordan-Wigner fermions is nonlocal, but
it is possible to obtain $\langle\sigma^x\rangle$ from the large-$r$
limit of the correlation function
$\langle\sigma^x_j\sigma^x_{j+r}\rangle$ \cite{pfeuty:1970a},
yielding
\begin{equation}
\langle\sigma^x\rangle =
\begin{cases}
0,\quad \lambda \le 1,\\
(1-\lambda^{-2})^{\frac{1}{8}}, \quad \lambda>1.
\end{cases}
\end{equation}
The transverse magnetisation $\langle\sigma^z\rangle$ is given by
the integral Eq.~(\ref{eq:trmag}) which reduces to an elliptic
integral for $\gamma=1$ and $\beta\rightarrow\infty$,
\begin{equation}
\langle\sigma^z\rangle = \frac{1}{\pi}\int_0^\pi d\phi
\frac{1+\lambda\cos\phi}{\sqrt{1+\lambda^2+2\lambda\cos\phi}}.
\end{equation}
Armed with knowledge of the appropriate correlation functions we can
now proceed to the calculation of the entanglement in the ground state of the
$XY$ and transverse Ising models.

\subsection{Ground state entanglement in the transverse Ising model}\label{ssec:gsent}

Given the modern understanding of entanglement as a physical resource
it makes sense to ask \emph{how much} entanglement there is in a given
multipartite state.  In order to answer this question the notion of an
\emph{entanglement measure} has been developed. A review of work on
entanglement measures may be found in~\cite{horodecki:2001a,
wootters:2001a, horodecki:2001b, nielsen:2001a}.

The study of entanglement measures is far from completely developed.
There is currently no consensus as to the best method to define an
entanglement measure for all possible multipartite states.  There are,
however, situations where there is an unambiguous way to construct
suitable measures.  It is these situations that we study in this
paper.

When a bipartite quantum system $AB$ is in a pure state there is
an essentially unique measure of the entanglement between the subsystems
$A$ and $B$ given by the \emph{von Neumann entropy} $S$
\cite{bennett:1996a, popescu:1997a, vidal:2000b, nielsen:2000b}.
The von Neumann entropy is calculated from the reduced density
matrix $\rho_A$ or $\rho_B$ according to the formula
\begin{equation}
S \equiv -\tr(\rho_A\log\rho_A)=-\tr(\rho_B\log\rho_B).
\end{equation}
When either subsystem $A$ or $B$ is a spin-$\frac12$ system, $S$ varies
from $0$ (product state) to $S=1$ (maximally entangled state).
For the ground state of the transverse Ising model we regard a single site
as subsystem $A$ and the rest of the lattice as subsystem $B$.

When a bipartite system $AB$ is in a mixed state there are a number of
proposals for measures of the entanglement in the state, including,
the \emph{entanglement of formation} \cite{bennett:1996a,
wootters:2001a}, the \emph{distillable entanglement}
\cite{bennett:1996a, bennett:1996c}, and the \emph{relative entropy of
entanglement} \cite{vedral:1997a, vedral:1998a}.  Each of these
measures have the property that, for pure states of $AB$, they reduce
to the von Neumann entropy.  The entanglement of formation
$\mathscr{F}(A:B)$ is the best understood of the mixed-state
entanglement measures.  For this reason, in this paper, we use the
entanglement of formation to measure the mixed-state entanglement in
the $XY$ model.

At the current time, there is no simple way to calculate the
entanglement of formation for mixed states of bipartite systems $AB$
where the dimension of $A$ or $B$ is three and above.  However, for
the case where both subsystems $A$ and $B$ are spin-$\frac12$
particles there exists a simple formula from which the entanglement of
formation can be calculated \cite{wootters:1998a}.  In this case the
entanglement of formation is given in terms of another entanglement
measure, the \emph{concurrence} $C$ \cite{hill:1997a, wootters:1998a,
wootters:2001a}.  The entanglement of formation varies monotonically
with the concurrence.

The entanglement $S$ between a single site and the rest of the lattice
represents the collective contibutions of the entanglement between the
given site and all other sites in the lattice.  Unfortunately the
single-site entanglement does not tell us how the entanglement is
shared out.  For example, $S=1$ could mean that the site in question
is maximally entangled with a neighbouring site, \emph{or}, entangled
with many sites.  In the transverse Ising model it appears that $S$ is
related to the onset of correlations in a fairly direct way (see
below), and to reflect this we speak of $S$ as `measuring' how
entangled the lattice is.

We should point out that this situation is by no means typical.  It is
quite common for the ground state of a condensed matter system to
possess strong nearest-neighbour entanglement and no long-range
correlations (see, for example, the models constructed by Affleck,
Kennedy, Lieb and Tasaki discussed in \cite{auerbach:1994a}).
Analysis of the entanglement in various AKLT models carried out by the
authors has shown that, in fact, the single-site entanglement is
\emph{constant} for all parameter values even though long-range
correlations develop and vanish.  The entanglement in these models
(and many other condensed matter systems) is, in general, not revealed
from knowledge of the single site density matrix.  What is really
needed --- but which has not yet been developed --- to study these
models is an entanglement measure which can take account of the way
entanglement is shared out.

\begin{figure}
\begin{center}
\includegraphics{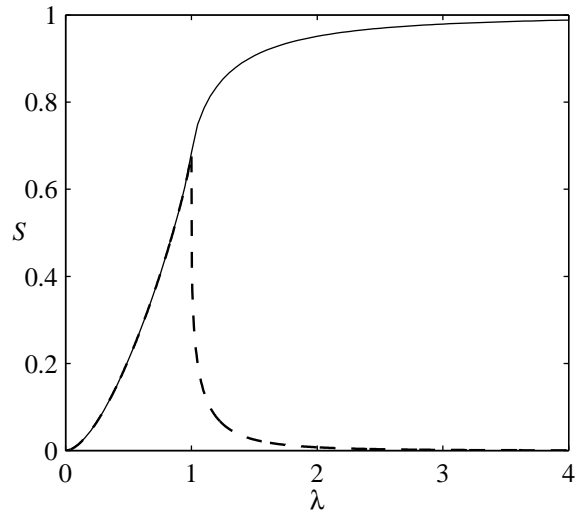}
\caption{Single-site entropy $S$ for the thermal ground state $\rho_0$ (solid)
and the single-site entanglement for the ground state $|0^+\rangle$
(dashed) of the transverse Ising
model}\label{fig:gse}
\end{center}
\end{figure}

At the critical point, $\lambda_c=1$, of the transverse Ising model
there is a fundamental transition in the structure of the ground
state. The correlation function $ \langle\sigma_i^\alpha\sigma^\beta_j
\rangle - \langle\sigma_i^\alpha \rangle \langle\sigma_j^\beta \rangle
$ decays polynomially as a function of separation at this point (the
dominant term has exponent $-\frac14$) while for all other values of
$\lambda$ this decay is exponential.  Interestingly, one could argue
that the correlation function itself actually constitutes an
entanglement measure for pure states as it transforms as a tensor
under local unitary operations and is zero for product states.  As
argued earlier, the change in the correlation function signals a
fundamental change in the entanglement present in the ground state.
This change is reflected in the single-site entanglement $S$ for the
ground state which appears in Fig.~\ref{fig:gse}.  The single-site
entanglement varies from zero at $\lambda=0$, where the ground state
is a product, to a maximum at the critical point $\lambda=1$.  As the
limit $\lambda\rightarrow\infty$ is approached $S$ also approaches
zero because the ground state again approaches product form.
The single-site von Neumann entropy for the thermal 
ground state of the
transverse Ising model is also shown in Fig.~\ref{fig:gse}.  Unlike the
ground state case, the entropy approaches unity in the limit
$\lambda\rightarrow\infty$.  This is because the thermal ground
state approaches an equal mixture of two pure states (the
eigenstates $|0^+\rangle$ and $|0^-\rangle$) in this limit.  The
single-site entropy is not measuring the entanglement content of
the thermal ground state in this limit, rather it is measuring the degree of
mixedness of the thermal ground state.

It is an intriguing fact that systems with quite different microscopic
dynamics may behave equivalently at criticality.  Further, their
behaviour depends only on the dimension of the system and the symmetry
of the order parameter.  The character of this behaviour is captured
by a small number of \emph{universal} quantities whose behaviour at
criticality is completely described in terms of a unique \emph{single}
number, a \emph{critical exponent}.  The equivalence of physically
different systems and their simple dependence on certain global
properties at criticality is known as \emph{universality}.  One of the
triumphs of twentieth century physics was the development of the
renormalization group (RG), which provided an explanation for the
emergence of universality in critical systems.

If we are to suppose that $S$ is a universal quantity which could be studied
via the RG then
we should be able to find a critical
exponent
for $S$.  In other words, near the
critical point we should be able to write something like
\begin{equation}\label{eq:entcrit}
S \propto |\lambda-\lambda_c|^\gamma,
\end{equation}
where $\gamma$ is the critical exponent for $S$.  
Unfortunately, this is
not possible.  As we describe below, 
the single-site entanglement is two-sided, so that \emph{two}
numbers are needed to specify $S$ like Eq.~(\ref{eq:entcrit}) near the
critical point, one for each
of the two ways of approaching $\lambda_c=1$.  In this way we see that the
single-site entanglement is \emph{not} a universal quantity.

The two-sided behaviour of the single-site entanglement 
arises because the single-site density matrix depends on both the 
magnetisation and the transverse magnetisation.  In the region near 
$\lambda \le 1$ 
only the transverse magnetisation is nonzero and 
the single-site entropy rises linearly.  At the critical point the 
magnetisation becomes nonzero and increases as $\lambda^{\frac18}$.  This 
becomes the dominant term in the expression for the single-site 
entanglement, and so the decay of the single-site entanglement is faster
than linear in the region near $\lambda >0$.  

If there exist universal quantities related to the entanglement in
critical quantum systems, then it is likely that they are derived from
entanglement measures that satisfy additional properties beyond the
set usually regarded as `essential' for an entanglement measure
(see, for example,~\cite{vedral:1998a, vidal:2000b}).  There are two
main reasons why we make this assertion.  The first arises from the
inability of the single-site entanglement to distinguish between
neighbouring and distributed entanglement.  In order to distinguish
between these differing scenarios, a good entanglement measure for
critical quantum systems should take account of how the entanglement
is shared out.  The second reason is that, as we argue below, the
single-site entanglement is not \emph{rescaleable}.  If a quantity is
to be renormalizable it is necessary that it be rescaleable.  That is,
it must be possible to collect degrees of freedom together, calculate
the collective value of the quantity, and then rescale (or
`renormalize') the collective value.  A renormalizable entanglement
measure should be rescaleable in this way.

We should be a little more precise in our definition of rescaleability
for entanglement measures.  Say we wish to calculate the bulk
entanglement of a block of spins $s_1, s_2,\ldots,s_m$ in a lattice
with the rest of the lattice, $L$.  If the entanglement measure
$\mathscr{G}$ (for example, $\mathscr{G}$ could be the entanglement of
formation) used to calculate this entanglement is to be rescaleable
then, in the very least, it must satisfy the \emph{extensivity
relation}
\begin{equation}\label{eq:extens}
\mathscr{G}(s_1, s_2,\ldots,s_m:L) \ge \mathscr{G}(s_1:L) + 
\cdots + \mathscr{G}(s_m:L).
\end{equation}
This inequality expresses the idea that the entanglement of a
collection of spins with the rest of the lattice should be \emph{at
least as great} as the sum of the entanglements of each spin with $L$.
If an entanglement measure does not satisfy the extensivity relation
Eq.~(\ref{eq:extens}) then it is not clear how to rescale the bulk
value of the entanglement.

Summarizing, the failure of the single-site entanglement to be
universal may be due to the facts that: (a) it does not distinguish
localized from distributed entanglement; and (b) it is not
rescaleable, in a sense that we can now make explicit.  To do this,
note first that it has previously been shown that the entanglement of
formation does not satisfy Eq.~(\ref{eq:extens}) \cite{coffman:2000a}.
If we regard the single-site entanglement $S$ as the entanglement of
formation $S=\mathscr{F}(s_1,L)$ between a single spin $s_1$ and the
rest of the lattice $L$ it seems unlikely that it will be a universal
quantity.  (There do exist other entanglement measures which reduce to
the von Neumann entropy for pure states
\cite{bennett:1996a,bennett:1996b,vedral:1997a,vedral:1998a}.  It is
an open question whether they satisfy Eq.~(\ref{eq:extens}).)  

There are indications \cite{coffman:2000a}, however, that the
\emph{square} of the concurrence \emph{is} extensive.  Perhaps a
suitable generalisation of the concurrence will turn out to be the
best quantity for studying universal properties of entanglement.
Evidence that this is the case has recently been obtained by Osterloh
\emph{et.\ al.} \cite{osterloh:2002a} where they found that a quantity
related to the concurrence \emph{is universal} for the transverse
Ising and $XY$ models.  It would be interesting to investigate this
behaviour and see if it arises because of the possible extensivity
properties of the concurrence.  Note, incidentally, that universal
behaviour in the concurrence does not necessarily imply universal
behaviour for the entanglement of formation, for the latter is only a
function of the former in the special case of a two-qubit system.

The determination of what entanglement is shared by two sites in the
lattice requires a measure of the two-party entanglement present in
mixed states.  We will henceforth use the concurrence $C$ to measure
the two-party mixed-state entanglement between two spins.  The
concurrence of two spin-$\frac12$ particles may be calculated from
their density matrix $\varrho$ via the formula
\begin{equation}
C(\varrho) = \max[0,\lambda_1-\lambda_2-\lambda_3-\lambda_4]
\end{equation}
where the $\lambda_i$ are the eigenvalues in decreasing order, of
the Hermitian matrix
$R\equiv\sqrt{\sqrt{\varrho}\tilde{\varrho}\sqrt{\varrho}}$, and
$\tilde{\varrho}=(\sigma^y\otimes\sigma^y)\varrho^*(\sigma^y\otimes\sigma^y)$.
The concurrence varies from $C=0$ for a
separable state to $C=1$ for a maximally entangled state
\cite{endnote61}.

The two-site density matrices for the ground state of the $XY$ model 
are difficult to calculate
when there is ground state degeneracy.  This is because
the magnetisation $\langle\sigma^x\rangle$ becomes nonzero as the
phase-flip symmetry is broken, and it becomes necessary
to include the correlation function $\langle\sigma^x_0\sigma^z_r\rangle$ in the
operator expansion Eq.~(\ref{eq:ope}).  The
$\langle\sigma^x_0\sigma^z_r\rangle$ correlation function is nonlocal when
expressed in terms of the Jordan-Wigner fermionic operators and there is no
simple way to derive it from other correlators.  As a result of this
difficulty we do not calculate the two-site density matrix for the ground
state, instead,
all two-site calculations are performed with respect to the thermal
ground state.  However, because the thermal
ground state for the transverse Ising model takes the special form
Eq.~(\ref{eq:gsens}), it is possible to place bounds on the entanglement
that can occur between
two sites in a degenerate ground state.

\begin{figure}
\begin{center}
\includegraphics{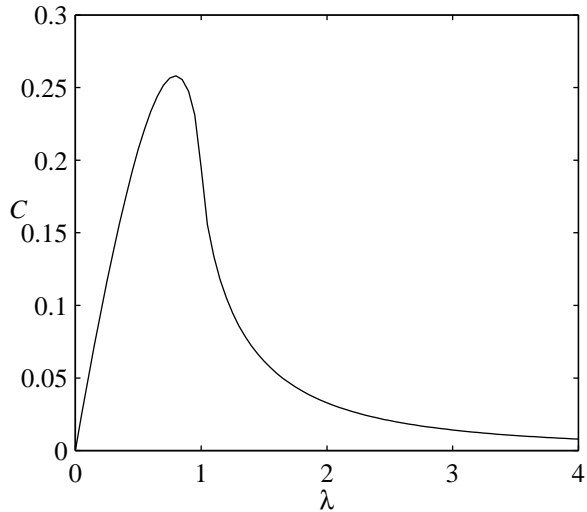}
\caption{Nearest-neighbour concurrence $C$ at zero temperature for the
transverse Ising
model}\label{fig:nnconc}
\end{center}
\end{figure}

\begin{figure}
\begin{center}
\includegraphics{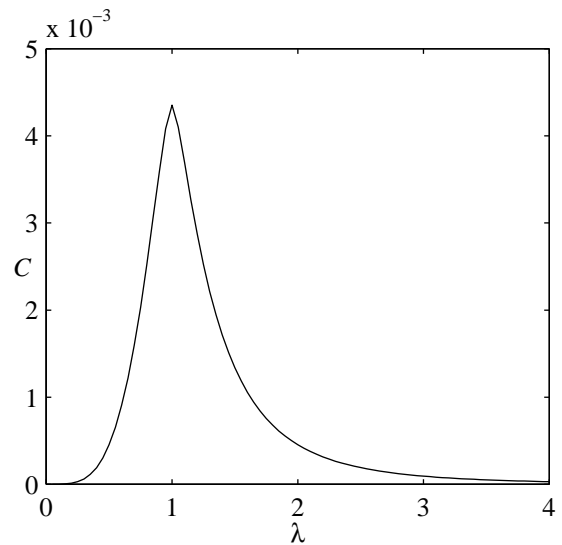}
\caption{Next-nearest-neighbour concurrence $C$ at zero temperature
for the transverse Ising
model}\label{fig:nnnconc}
\end{center}
\end{figure}

The entanglement between pairs of sites for the thermal ground state
of the transverse Ising model shares many of the same features of the
single-site entanglement.  The entanglement, as measured by the
concurrence, between neighbouring sites and next-nearest neighbouring
sites is shown in Fig.~\ref{fig:nnconc} and Fig.~\ref{fig:nnnconc}
respectively.  All other pairs have zero two-party entanglement
because the correlation functions drop below the threshold for a
positive concurrence.  In both cases the entanglement rises from zero
in the limits $\lambda=0$ and $\lambda\rightarrow\infty$ to a maximum
value near the critical point $\lambda=1$.  When $\lambda\le1$ the
ground state coincides with the thermal ground state so that the
two-site entanglement results are the same in this case.  Note that
the maximum does not occur {\em exactly} at the critical point
$\lambda = 1$.  At first site this may appear to contradict our
earlier conjecture that we expect entanglement to be the greatest at
the critical point.  In fact, as explained in
Sec.~\ref{ssec:entshare}, the reason for this is that the results
here are for two-site entanglement, and are not inconsistent with the
conjecture that the {\em total} entanglement in the lattice is a
maximum at the critical point.

It is interesting to see what effect the ground state degeneracy has
on the two-site entanglement in the ground state.  As
mentioned, it is not possible to study the two-site entanglement for
$\lambda>1$.
Despite this difficulty, for $\lambda$ above the critical value,
we can place a lower bound
on the two-site entanglement in a degenerate ground state.
This may be
achieved by observing that the
concurrence measure $C$ is \emph{convex} \cite{wootters:1998a}, 
which means that
\begin{equation}
C\left(\sum_{i=1}^n p_i\rho_i\right) \le \sum_{i=1}^n p_iC(\rho_i),
\end{equation}
where $p_i$ is any probability distribution and $\rho_i$ a set of two-site
density matrices.  If we apply this inequality to the thermal ground state,
Eq.~(\ref{eq:gsens}), we obtain $C(\rho_{0r}) \le
\frac12C(\tr_{\widehat{0r}}(|0^+\rangle\langle0^+|)) +
\frac12C(\tr_{\widehat{0r}}(|0^-\rangle\langle0^-|))$.  The global phase
flip is a local unitary operation, so that the concurrence of each term in
the RHS of the inequality is the same, that is 
\begin{equation}\label{eq:cineq}
C(\rho_{0r}) \le C(\tr_{\widehat{0r}}(|0^+\rangle\langle0^+|)).
\end{equation}
In this way we see that the two-party entanglement in the ground state is at 
least as large as the two-party entanglement in the thermal ground state.

\subsection{Critical quantum systems and the constraints of shared
entanglement}\label{ssec:entshare}

The maximum value of the concurrence between neighbouring sites does
not occur at the critical point.  This seemingly contradicts the idea
that the strength of the correlations is proportional to the
entanglement, and that therefore the entanglement should be maximal at
the critical point.  However, as we will discuss in this subsection,
there are reasons based on the properties of shared entanglement to
expect that this maximum should occur away from the critical point.

It is well known that there are limitations to the amount of
entanglement that may be distributed amongst three or more subsystems
\cite{bruss:1999a, coffman:2000a, koashi:2000a, dur:2000a, wootters:2000a, oconnor:2001a,
 dennison:2001a}. This class of problem, that is, the determination of
how much two-party entanglement can be distributed amongst a given
number of parties, is known as an \emph{entanglement sharing}
problem. The simplest example of this is the situation of three
parties $A$, $B$ and $C$.  If $A$ is maximally entangled with $B$ then
it is not possible for $A$ and $C$ or $B$ and $C$ to share any
two-party entanglement. Entanglement sharing is relevant to the
quantum phase transition in the transverse Ising model as it provides
a fundamental bound on the amount of entanglement that may be
distributed amongst the sites.  The existence of such a bound means
that as the overall entanglement in the lattice is increased, some
sites become pairwise more disentangled.  An example where this occurs
is in a system approaching a critical point.

As the critical point is approached in the transverse Ising model the
correlation length begins to increase.  What occurs physically is that
each site develops entanglement with its neighbouring sites.  When the
system gets closer to the critical point each site begins to develop
entanglement with its next-nearest neighbours and so on.  When the
system is not at the critical point the entanglement between a single
site and the rest of the lattice is localised within some region
because the correlations are exponentially damped for large enough
separation.  At the critical point this is no longer the case; there
are appreciable correlations between a single site and every other
site.  However, the entanglement associated with this correlation must
be distributed in such a way that it satisfies the constraints of
entanglement sharing.  We conjecture that the ground state at the
critical point actually \emph{saturates} the bounds of entanglement
sharing, so that it is \emph{maximally entangled} in this sense.  If
this conjecture is correct, this would explain why the entanglement
between neighbouring sites is not maximum at the critical point.
Initially, as $\lambda$ is increased, the entanglement between
neighbouring sites increases first.  When the system reaches
criticality the entanglement is distributed to more remote pairs.  If
the ground state saturates the bounds of entanglement sharing this
would have to occur at the expense of the two-party entanglement
previously established between pairs of sites that are close.

In the light of this interpretation it is interesting to compare the
entanglement calculations for the transverse Ising model at
criticality to the lattice calculations of Wootters and O'Connor
\cite{wootters:2000a,oconnor:2001a}. In the critical case
$\lambda_c=1$ the correlation functions for the transverse Ising model
are known explicitly as functions of $r$ \cite{pfeuty:1970a}
\begin{align} \langle \sigma_0^x\sigma_r^x\rangle &=
\left(\frac{2}{\pi}\right)^r2^{2r(r-1)}\frac{H(r)^4}{H(2r)},\\ \langle
\sigma_0^y\sigma_r^y\rangle &= -\frac{\langle
\sigma_0^x\sigma_r^x\rangle}{4r^2-1},\\ \langle
\sigma_0^z\sigma_r^z\rangle &= \frac{4}{\pi}\frac{1}{4r^2-1},\\
\langle\sigma^z\rangle &= \frac{2}{\pi}, \end{align} where
$H(r)=1^{r-1}2^{r-2}\dots(r-1)$. The concurrence at the critical point
is nonzero for both $r=1$ and $r=2$ where it is given by,
respectively, $0.1946$ and $0.0044$.  These values should be compared
with the values obtained by O'Connor and Wootters in their study
\cite{oconnor:2001a} of the concurrence in chains and rings of
qubits. They maximised the entanglement between nearest neighbours of
a translationally invariant ring of spin-$\frac12$ degrees of
freedom. Wootters and O'Connor were attempting to saturate the bounds
of entanglement sharing by maximising the entanglement of nearest
neighbours subject to the symmetry of translational invariance. They
found a maximal nearest-neighbour concurrence value of $0.4345$ for an
infinite ring, which is greater than the critical value for the
transverse Ising model.  This result alone does not imply that the
critical transverse Ising model is less entangled than the ring
considered in \cite{wootters:2000a,oconnor:2001a}, indeed, if the
conjecture made in the previous paragraph is true then the ring would
be much less entangled than the critical transverse Ising model.  The
reasoning for this is that the critical transverse Ising model is
conjectured to maximise the entanglement between all pairs subject to
translational invariance while the chains and rings of Wootters and
O'Connor only maximise entanglement between nearest neighbours. One
means of determining whether this is the case would be to calculate
the correlation function for the ring.  On the basis of the arguments
made in this study, we expect that the correlations will decay
exponentially with separation for the ring.

\begin{figure}
\begin{center}
\includegraphics{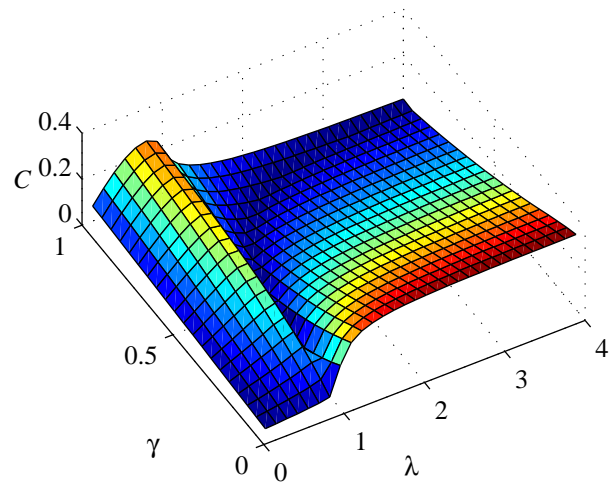}
\caption{Nearest-neighbour concurrence $C$ at zero temperature
for the $XY$
model}\label{fig:nnxy}
\end{center}
\end{figure}

\begin{figure}
\begin{center}
\includegraphics{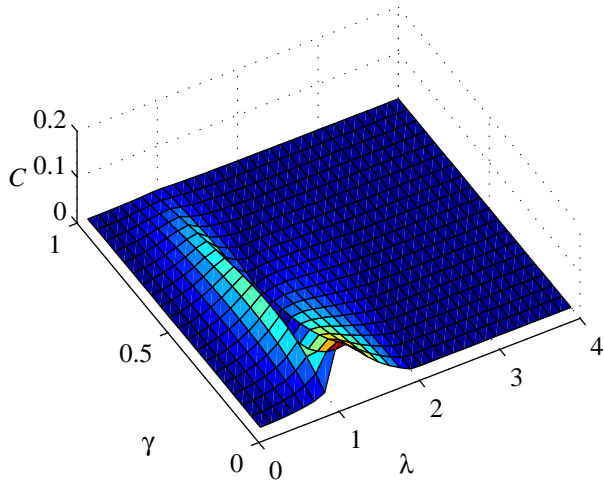}
\caption{Next-nearest-neighbour concurrence $C$ at zero temperature
for the $XY$
model}\label{fig:nnnxy}
\end{center}
\end{figure}

The entanglement in the thermal ground state of the general $XY$ model
may be calculated simply, following the method outlined in
Sec.~\ref{sec:exsol}. Following Barouch \cite{barouch:1971a}, which is
where the correlation functions Eq.~(\ref{eq:corr1}),
Eq.~(\ref{eq:corr2}), and Eq.~(\ref{eq:corr3}) were calculated, only
the region $0\le\gamma\le1$ is considered here. The concurrence
between nearest-neighbour and next-nearest neighbour sites is shown in
Fig.~\ref{fig:nnxy} and Fig.~\ref{fig:nnnxy} respectively. The
concurrences are a complicated function of the parameters, reflecting
the competition between the various different noncommuting terms in
the Hamiltonian as the parameters are varied.

The completely isotropic limit, $\gamma=0$, is the most interesting
parameter region besides the transverse Ising model.  Direct
calculation along the lines already presented shows that two-party
entanglement exists between all pairs for all separations at this
point.  Wootters~\cite{wootters:2002a} has made a study of the
correlations in one- and two-dimensional lattices and he has found
interesting connections between the two-party correlations in the
isotropic $XY$ model and the bounds of entanglement sharing.  Further
investigations along these lines could provide evidence that critical
quantum lattice systems are maximally entangled in the sense of
entanglement sharing.

\section{Thermal entanglement in the transverse Ising model}\label{sec:2sent}

In this section we discuss the entanglement present in the thermal state
of the transverse Ising model.  We find that the largest amount of 
entanglement is present in the parameter region close to the critical
point.  This region is found to correspond with the \emph{quantum critical}
region introduced by Sachdev (\cite{sachdev:1999a}, pg.~58).  We also find parameter
values for which the entanglement \emph{increases} as the temperature is
increased.  Finally, we discuss the persistence of quantum effects in the
thermal state as the temperature is increased.

\begin{figure}
\begin{center}
\includegraphics{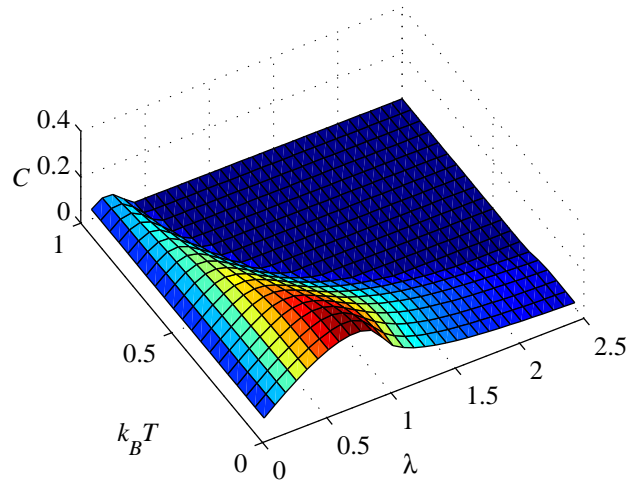}
\caption{Nearest-neighbour concurrence $C$ at nonzero temperature
for the transverse Ising
model}\label{fig:nnct}
\end{center}
\end{figure}

\begin{figure}
\begin{center}
\includegraphics{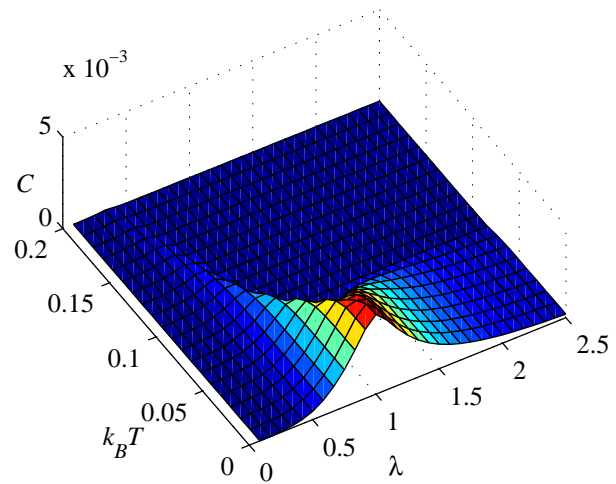}
\caption{Next-nearest-neighbour concurrence $C$ at nonzero temperature
for the transverse Ising
model}\label{fig:nnnct}
\end{center}
\end{figure}

It is desirable to determine when a condensed-matter system will
behave quantum-mechanically.  This is particularly important because the
validity of various ansatz methods depends on whether they take account of
possible quantum effects. 
When a system is in its ground state
quantum effects will certainly be important, as evidenced by the
quantum phase transition in the $XY$ model. The zero-temperature
calculations of the last section represent a highly idealised
situation, however, and it is unclear whether they have any relevance to
the system at nonzero temperature.  
It turns out that the properties of a quantum
system for low temperatures are strongly influenced by nearby (in
parameter space) quantum critical points \cite{sondhi:1997a, sachdev:1999a}.
It is tempting to attribute the effect of nearby critical points to
persistent mixed-state entanglement in the thermal state. 
In order to investigate this, we calculate the two-party entanglement
present at a nonzero temperature $T$.  

The two-site density matrices constructed
in Sec.~\ref{sec:exsol} are valid for all temperatures.  Using these
matrices it is possible to study the purely two-party entanglement present
at thermal equilibrium because
the concurrence
measure of entanglement can be applied to arbitrary mixed states.
The regions where there
is appreciable two-party entanglement give at least a partial indication
of where quantum effects may be important.
We again emphasise the transverse Ising model for this section.  The
influence the critical point has on the entanglement structure at nonzero
temperatures is particularly clear for this model.

The entanglement between nearest-neighbour and next-nearest
neighbour sites in the Ising model at nonzero temperature
appears in Fig.~\ref{fig:nnct} and Fig.~\ref{fig:nnnct}
respectively.  The entanglement is nonzero only in a certain
region in the $k_BT-\lambda$ plane.  It is in this region that
quantum effects are likely to dominate the behaviour of the system.  The
entanglement is largest in the vicinity of the critical point
$\lambda=1$, $k_BT=0$. This region corresponds, approximately, to the
\emph{quantum critical} regime identified by Sachdev
\cite{sachdev:1999a}.  Sachdev found, by using a very different
argument, that quantum effects would be important in this regime.
The correspondence of these two regions provides evidence that the
entanglement content plays an important role in the emergence of
quantum behaviour in naturally occurring quantum systems.

There are two notable features of the two-site thermal entanglement
results. The first feature is that, for certain values of $\lambda$,
the two-site entanglement can increase as the temperature is increased
(eg.\ $\lambda=1.4$, Fig.~\ref{fig:nnct}).  This effect has previously
been observed in finite-size calculations
\cite{nielsen:1998a, arnesen:2001a} for the Heisenberg model.  The
occurrence here of the same effect implies that it is not an artifact
of the truncation of a lattice. The second feature is the existence of
appreciable entanglement in the system for temperatures $k_BT$ above
the ground state energy gap $\Delta$. It has been argued
\cite{sondhi:1997a} that quantum systems behave \emph{classically}
when the temperature exceeds all relevant frequencies. For the
transverse Ising model the only relevant frequency is given by the
ground state energy gap $\Delta\equiv\hbar\omega$. The presence of
entanglement in the system for temperatures above the energy gap
indicates that quantum effects may persist past the point where they
are usually expected to disappear.

A comparison should be made between the results obtained here and
the numerical calculations of concurrence in the Ising model on a
finite number of sites \cite{gunlycke:2001a}.  The calculations
that were performed in \cite{gunlycke:2001a} were implemented on a
maximum of $7$ sites.  The concurrence between nearest neighbours
obtained by Gunlycke \emph{et al}.\ (Figs.~$2$ and $5$ of
\cite{gunlycke:2001a}) is in qualitative agreement with the
results obtained here.  However, as there is no phase transition
for the finite size Ising model the dominance of the critical
point was not as sharp in the calculations of
\cite{gunlycke:2001a}.

\section{Summary and Future directions}\label{sec:conc}

The one- and two-party entanglement present in the ground and thermal
states of the $XY$ model has been calculated.  It should be stressed
that the calculations in this study are analytic and, furthermore,
they are for the thermodynamic limit of a quantum lattice system.

We have argued that the critical point of a quantum lattice system
corresponds to the situation where the lattice is maximally entangled.
Evidence for this conjecture was found in the single-site entanglement
results for the ground state of the transverse Ising model.  We have
also argued that the constraints of shared entanglement are important
for critical quantum systems, and we have found possible evidence of
such constraints playing a role in the two-party entanglement results
for the transverse Ising model.  The entanglement present at thermal
equilibrium was also studied, and an approximate correspondence
between the quantum critical regime identified by Sachdev and the
regions where the two-party entanglement is nonzero was found.
Parameter values where the entanglement increases as the temperature
is increased were also found.

We have focused on the transverse Ising model throughout this study,
although the calculations presented also cover the $XY$ model. The
transverse Ising model is interesting because it is the simplest
system to exhibit a quantum phase transition, and it is relatively
easy to identify the structure of the entanglement present in this
system.  The importance of the critical point in this system is also
particularly clear.  The $XY$ model has many parameter regimes where
it behaves differently, so it is very likely that more interesting
phenomena may be found in other parameter regions.

Entanglement calculations in this study have been restricted to
time-independent scenarios.  However, the dynamic correlation
functions have been calculated for the Ising and $XY$ models for
certain values of $\lambda$.  It is possible and may be interesting to
calculate the time evolution of the entanglement in these models and
thus identify truly quantum dynamics.

The calculations in this study are intended as a point of reference
for the development of an understanding of the entanglement in
critical quantum systems.  Rather frustratingly, the present
incomplete understanding of entanglement measures has prevented us
from performing many of the calculations we would like to do in order
to check the many conjectures made in this paper.  Further progress on
the general quantitative theory of entanglement should enable these
conjectures to be checked in the future.  We believe that entanglement
plays a central role in the emergence of long-range correlations at
the critical point of such systems, and that a fruitful interplay
between the theory of entanglement and critical quantum phenomena may
result from further study.  In particular, it would be interesting to
make {\em universal} statements about the character of entanglement at
the critical point, and to examine whether the constraints of
entanglement sharing impose physical limitations on the behaviour that
can occur in such a system.

\emph{Note added:} As this paper was nearing completion we learnt of
related work done independently by Osterloh \emph{et.\ al.}
\cite{osterloh:2002a}.

\begin{acknowledgments}
We would like to thank 
Dorit Aharonov, Nick Bonesteel, John Preskill and Bill Wootters for many 
stimulating and encouraging discussions about entanglement and 
phase transitions.  We would also like to thank Jennifer Dodd, Alexei
Gilchrist, Ross McKenzie and Howard Wiseman for their helpful comments
on the manuscript.
This work has been funded, in part, by an Australian Postgraduate Award
to TJO.
\end{acknowledgments}

\end{document}